\documentclass[times,trackchanges,twocolumn]{aastex7}
\usepackage{amsmath}
\graphicspath{{./}{figures/}}
\usepackage{txfonts}

\begin{document}

\title{A Radially Resolved Magnetic Field Threading the Disk of TW~Hya}

\author[0000-0003-1534-5186]{Richard Teague}
\affiliation{Department of Earth, Atmospheric, and Planetary Sciences, Massachusetts Institute of Technology, Cambridge, MA 02139, USA}
\email[show]{rteague@mit.edu}

\author[0000-0001-8975-9926]{Boy Lankhaar}
\affiliation{Department of Space, Earth and Environment, Chalmers University of Technology, Onsala Space Observatory, 439 92 Onsala, Sweden}
\affiliation{Leiden Observatory, Leiden University, Post Office Box 9513, 2300 RA Leiden, Netherlands}
\affiliation{Institute of Theoretical Astrophysics, University of Oslo, P.O. Box 1029, Blindern, 0315 Oslo, Norway}
\email{rteague@mit.edu}

\author[0000-0003-2253-2270]{Sean M. Andrews}
\affiliation{Center for Astrophysics \textbar\ Harvard \& Smithsonian, Cambridge, MA 02138, USA}
\email{rteague@mit.edu}

\author[0000-0001-8642-1786]{Chunhua Qi}
\affiliation{Center for Astrophysics \textbar\ Harvard \& Smithsonian, Cambridge, MA 02138, USA}
\affiliation{Institute for Astrophysical Research, Boston University, 725 Commonwealth Avenue, Boston, MA 02215, USA}
\email{rteague@mit.edu}

\author{Roger R. Fu}
\affiliation{Department of Earth and Planetary Sciences, Harvard University, Cambridge, MA, USA}
\email{rogerfu@fas.harvard.edu}

\author[0000-0003-1526-7587]{David J. Wilner} 
\affiliation{Center for Astrophysics \textbar\ Harvard \& Smithsonian, Cambridge, MA 02138, USA}
\email{rteague@mit.edu}

\author[0000-0001-5243-241X]{John B. Biersteker}
\affiliation{Department of Earth, Atmospheric, and Planetary Sciences, Massachusetts Institute of Technology, Cambridge, MA 02139, USA}
\email{rteague@mit.edu}

\author[0000-0002-5758-150X]{Joan R. Najita}
\affiliation{NSF NOIRLab, 950 N. Cherry Avenue, Tucson, AZ 85719, USA}
\email{rteague@mit.edu}

\begin{abstract}
We present a new approach to detecting and characterizing a magnetic field in protoplanetary disks through the differential broadening of unpolarized molecular emission from CN. To demonstrate this technique, we apply it to new ALMA observations of the full complement of hyperfine components from the $N=1-0$ transition, achieving a spatial and spectral resolution of ${\approx}0\farcs5$ and $80~{\rm m\,s^{-1}}$, respectively. By fitting a model that incorporates the velocity structure of the disk, the potential non-LTE excitation of the molecule, and the Zeeman effect, we recover a radially resolved magnetic field with a strength of ${\sim}10~{\rm mG}$ between 60 and 120~au. The morphology of the field is also inferred through azimuthal variations in the line broadening, revealing a predominantly poloidal field at 60~au, sharply transitioning to one within the disk plane outside of the gap at 82~au. The signal-to-noise ratio of the data meant that the planar component was unable to be decomposed into toroidal and radial components. Lower limits on the local gas density ($n({\rm H_2}) \gtrsim 10^8~{\rm cm^{-3}}$) from the excitation analysis of the CN emission correspond to a lower limit between 0.1 and 0.01 for the plasma $\beta$.
\end{abstract}


\section{Introduction}
\label{sec:introduction}

Planets form in protoplanetary disks, the remnant gas and dust from the star formation process that settles around the newly formed star. A variety of mechanisms have been proposed to account for the evolution of these systems, facilitating the redistribution of angular momentum, the concentration of gas and dust into regions favorable for planet formation, and the eventual dissipation of the disk \citep{Morbidelli_ea_2024}. High angular resolution observations have revealed large-scale perturbations in both the gas and dust of these sources demonstrating their highly active nature \citep[see, for example,][and references therein]{Andrews_2020, Bae_ea_2023}.

While purely hydrodynamical (HD) instabilities have shown promise in accounting for this evolution, the highly specific conditions required for such instabilities to grow and sustain themselves argue against these being the dominant mechanisms moderating the evolutionary processes \citep{Lesur_ea_2023}. Instead, magneto-hydrodynamical (MHD) instabilities appear to be less sensitive to the range of disk properties if the disk is threaded by a magnetic field making them a more attractive mechanism to explain disk evolution. However, while magnetic fields are commonplace on larger scales earlier in the star formation process \citep{Pattle_ea_2023} and have been indirectly inferred to be present during the formation of our solar system \citep{Weiss_ea_2021}, the detection and characterization of magnetic fields threading protoplanetary disks remains a significant observational challenge. Without robust evidence of the presence of magnetic fields threading a protoplanetary disk and a constraint on their morphology, the importance of MHD processes for the evolution of protoplanetary disks remains unconstrained. 

Routinely used to trace magnetic fields in star forming regions, continuum polarization detected observed in multiple sources is mostly due to grain alignment mechanisms unrelated to magnetic fields \citep{Tazaki_ea_2017} or the self-scattered of emission \citep{Kataoka_ea_2015}. However, there is one recent report of magnetically aligned grains in HD~142527 \citep{Ohashi_ea_2025}, where the relative wavelength-independence of the polarization pattern suggested magnetic alignment by a field of ${\sim}~0.3$~{\rm mG} over a select range of azimuths in the disk.

\begin{figure*}
    \includegraphics[width=0.9875\textwidth]{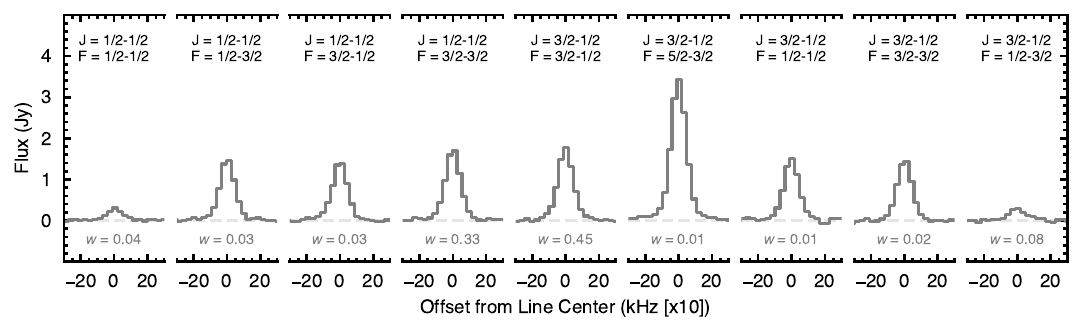}
    \caption{Observations of the nine hyperfine components of the CN $N=1-0$ transition from the TW~Hya disk used in this paper. The flux densities are calculated by integrating a circular mask with a radius of 4\arcsec. Each spectrum shows $\pm$ 300~kHz ($\approx 800~{\rm m\,s^{-1}}$) around the line center. The $w$ value under each line denotes the relative sensitivity of the component to Zeeman broadening. Note the sensitivity of the line to magnetic fields is uncorrelated with the intensity.}
    \label{fig:flux_density}
\end{figure*}

Searches for polarized molecular line emission have also been conducted. Both \citet{Stephens_ea_2020} and \citet{Teague_ea_2021} report the detection of linearly polarized CO emission in HD 142527 and IM~Lup, and TW~Hya, respectively, suggesting that the emission is due to the Goldreich-Kylafis effect \citep[GK;][]{Goldreich_Kylafis_1982}. While the detection itself signifies the presence of a magnetic field, the GK effect is insensitive to the magnetic field strength (above some negligible lower limit) and the averaging and stacking of the data required to pull out the detections removes any sensitivity to the field morphology. 

Attempts to detect the circular polarization of molecular emission due to the Zeeman effect were also carried out, however they yielded only upper limits in both TW~Hya and AS~209 \citep{Vlemmings_ea_2019, Harrison_ea_2021}. Similar to the linearly polarized emission, the unresolved highly structured emission morphology \citep{Mazzei_ea_2020} and cancellation of the polarization signal from the back side of the disk \citep{Lankhaar_Teague_2023} likely contributed to the non-detections. The high angular resolution and deep observations that these works suggest are necessary to detect circular polarization results in prohibitively long programs which would be exceedingly challenging to schedule. Alternative, less costly methodologies are needed.

In this paper, we present and apply such a technique to CN emission from the disk around TW~Hya. The detection of magnetic fields solely through the broadening of unpolarized molecular emission due to Zeeman splitting bypasses the need for polarized emission and benefits from the improved schedulability of low frequency programs. We describe the observations and data reduction in Section~\ref{sec:observations}, the modeling methodology in Section~\ref{sec:methodology} and results in Section~\ref{sec:results}. A discussion of these results and their implications are provided in Section~\ref{sec:discussion}, with Section~\ref{sec:conclusions} providing a summary of the findings.

\section{Observations}
\label{sec:observations}

The data from project 2022.1.00840.S (P.I. Teague) comprise ten execution blocks (EBs) observed between 7 January 2023 and 12 May 2023, using between 41 and 44 antennas, for a total on-source integration time of 670 minutes. Projected baselines ranged from 15.3 to 2516.8~m, and the maximum recoverable scale (MRS) was 12.2\arcsec, with a typical precipitable water vapor (PWV) range of 1.4-5.5~mm. Three spectral windows covered multiple CN $N=1-0$ transitions from 113.123 to 113.520~GHz, with a spectral resolution of 80~m\,s$^{-1}$. 

All individual EBs were first re-aligned to a common disk center, determined based on a Gaussian fit to the continuum emission, using the 
\texttt{phaseshift} and \texttt{fixplanets} tasks. We then performed three rounds of phase-only self-calibration on the short-baseline (SB) data with solution intervals (\texttt{solint}) of \texttt{inf}, 360\,s, and 90\,s), followed by one round of amplitude self-calibration with \texttt{solint}~=~\texttt{inf}. These calibrated SB data were concatenated with the long-baseline (LB) observations, after which two rounds of phase-only self-calibration (\texttt{solint}~=~\texttt{inf}, 360\,s) and one round of ampltitude self-calibration (\texttt{solint}~=~\texttt{inf}) were applied. 

This calibration yielded an improvement of approximately 4$\times$ in the peak continuum SNR. The final calibration tables were then applied to the aligned line visibilities. 

We adopt a $80~{\rm m\,s^{-1}}$ channel spacing, the FWHM spectral resolution of the data, and a \texttt{robust} value of 1 for the final imaging run, CLEANing down to a threshold of $3\sigma$. These parameters result in a median beam size of $0\farcs53 \times 0\farcs42$ with a position angle of $105\degr$, and a per channel sensitivity of $1.8~{\rm mJy~beam^{-1}}$ (0.76~K).

Flux densities for all nine hyperfine components are shown in Figure~\ref{fig:flux_density} which were calculated by spatially integrating a circulate aperture with a radius of 4\arcsec{} (240~au) in each channel using \texttt{GoFish} \citep{GoFish}. Below each line the $w$ value represents the relative sensitivity to Zeeman splitting, measured as the product of $Z$ and $\bar{Q}$, where $Z$ is the line-specific Zeeman coefficient and $\bar{Q}$ quantifies the contribution of the intra-group broadening to the total Zeeman broadening, as defined in \citet{Lankhaar_Teague_2023}. Note that the brightest components do not necessarily correspond to the components most sensitive to the Zeeman splitting. The total integrated flux for the $N=1-0$ transition is $4.38 \pm 0.04~{\rm Jy~km\,s^{-1}}$ with integrated fluxes for individual hyperfine components given in Table~\ref{tab:observations}.

\section{Methodology}
\label{sec:methodology}

In this section we describe the modeling methodology employed to infer the magnetic field strength and morphology. This builds on excitation analyses described in many previous works, such as \citet{Teague_ea_2016} and \citet{Teague_ea_2018c}, but with the addition of the Zeeman splitting of the lines.

\subsection{Excitation Analysis}
\label{sec:methodology:excitation}

At its core, this approach models the observed spectrum using the \texttt{SpectralRadex} \citep{SpectralRadex} implementation of \texttt{RADEX} \citep{vanderTak_ea_2007} with the collision rates from \citet{Kaligina_Lique_2015} to calculate the optical depth of each hyperfine component given a kinetic temperature, $T_{\rm kin}$, a column density of CN, $N({\rm CN})$, a local collider density, dominated by H$_2$ molecules, $n({\rm H_2})$, and non-thermal contribution to the line width, $\Delta V_{\rm nt}$. This model makes the implicit assumption that the emitting region can be well represented by a geometrically thin, isothermal slab. For the face-on case of TW~Hya, where no vertical extent can be observed, this has been shown to be an appropriate assumption and has been successfully used to reproduce a range of spectral observations \citep[e.g.,][]{Teague_ea_2018c, Romero-Mirza_ea_2023}.

Importantly, this approach accounts for both LTE and non-LTE solutions \citep[critical for CN as][suggest there may be subtle non-LTE effects present for CN emission from TW~Hya]{Teague_Loomis_2020}. Following \citet{Cataldi_ea_2021}, we account for any systematic broadening, such as instrumental broadening or due to unresolved Keplerian shear, by convolving the resulting spectrum with a Gaussian kernel with a FWHM that was allowed to vary, ${\rm FWHM_{conv}}$. To differentiate between non-thermal broadening (which would affect the optical depth profile) and instrumental broadening (which acts on the intensity profile) we include a Gaussian prior on $\Delta V_{\rm nt}$ centered on zero with a standard deviation of $0.08~c_s$ following the upper limits reported in \citet{Flaherty_ea_2018}.

A single pixel does not contain sufficient signal to robustly identify the influence of a magnetic fields on the line profile.  So instead, we simultaneously fit annuli of spatially independent pixels (i.e., pixels that are separated by at least a beam FWHM). The assumption here is that the disk is azimuthally symmetric such that the line profiles vary only in their centers due to the projection of the disk velocity structure. Previous works, such as \citet{Teague_ea_2016}, aligned and stacked these spectra prior to the fitting, following the now standard `shift-and-stack' approach described in \citet{Yen_ea_2016}. In order to incorporate the uncertainties associated with this process, we opt to instead shift our model spectra to the appropriate reference frame through
\begin{equation}
    v_0(r,\,\phi) = v_{\phi,\,\rm proj}(r,\,\phi) + v_{r,\,\rm proj}(r,\,\phi) + v_{z,\,\rm proj}(r,\,\phi) + v_{\rm LSR}
\end{equation}
where $v_{\phi,\,\rm proj}$, $v_{r,\,\rm proj}$ and $v_{z,\,\rm proj}$ are the azimuthally symmetric rotational, radial, and vertical velocity fields projected along the line of sight. These can be decomposed into their disk-frame components as
\begin{align}
    v_{\phi,\,\rm proj}(r,\,\phi) &= +v_{\phi}(r) \cos(\phi)\sin(|i|),\\
    v_{r,\,\rm proj}(r,\,\phi) &= -v_{r}(r) \sin(\phi)\sin(i)\\
    v_{z,\,\rm proj}(r,\,\phi) &= -v_{z}(r) \cos(i)
\end{align}
where $\phi$ is the polar angle in the disk frame and $i$ is the disk inclination. Note that the sign of $i$ is chosen here to encode the rotation direction of the disk, with positive values indicating clockwise rotation on the sky and negative values indicating counter-clockwise rotation. Assuming that the large spiral arm found in TW~Hya is trailing \citep{Teague_ea_2022} would imply that the disk is rotating in a clockwise direction and thus the $i$ in this parameterization is positive. Following \citet{Teague_ea_2022}, a source inclination and position angle of $+5\fdg8$ and $151\fdg6$ were adopted and used to calculate each annulus.

\subsection{Zeeman Splitting}
\label{sec:methodology:zeeman}

\begin{figure*}
    \centering
    \includegraphics[scale=1]{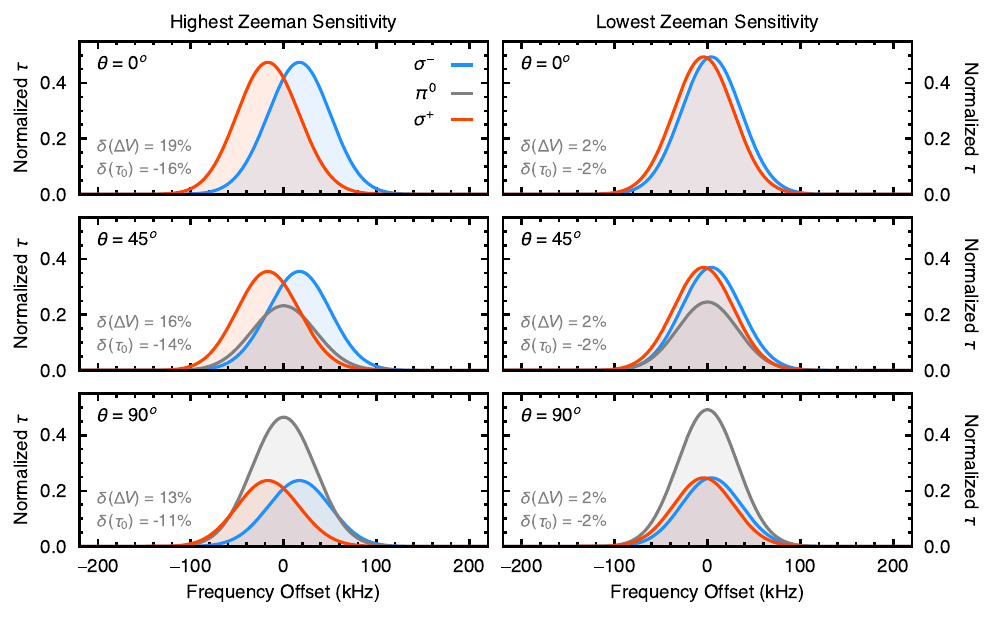}
    \caption{How the line strengths and positions of the three magnetic subtransition groups ($\pi^0, \sigma^+, \sigma^-$) vary when differing the orientation of a 15~mG magnetic field relative to the line of sight described by $\theta$ for the most sensitive hyperfine component, $N=1-0$ $J=3/1-1/2$ $F=3/2-1/2$, left, and the least sensitive component, $N=1-0$ $J=3/1-1/2$ $F=5/2-3/2$, right. Magnetic subtransition groups are influenced by a magnetic field, broadening the line profile an amount proportional to the sensitivity of the component to Zeeman broadening. At the 15~mG assumed here, the resulting line profile, the sum of the three components, is still close to a Gaussian profile. The gray annotations in the lower left of each panel show the relative change in line width and optical depth at the line center relative to an un-split transition. For the most sensitive component the changing field orientation leads to a different level of broadening and a corresponding change in the peak optical depth while for the least sensitive component there is only an imperceptible change with changing field orientation.}
    \label{fig:zeeman_splitting_demonstration}
\end{figure*}

Rather than adopt a Gaussian optical depth profile for each hyperfine component, we implement the Zeeman splitting model presented in \citet{Lankhaar_Teague_2023}. We represent the line profile of the CN transitions as a sum, weighted by the relative line strength, over all allowed transitions between the magnetic substates of the hyperfine levels. The energies associated with the magnetic substates are shifted with respect to the zero-field energy in proportion to the magnetic field strength, $|\mathbf{B}|$, the level-specific $g$-factor, the statistical weight of the level, and the associated magnetic quantum number $m$. The adopted molecular parameters are described in Table~\ref{tab:observations}. Figure~\ref{fig:zeeman_splitting_demonstration} demonstrates how the magnetic subtransition groups are influenced by a magnetic field, resulting in a broadened line profile. Note that at field strengths larger than the 15~mG used for Figure~\ref{fig:zeeman_splitting_demonstration}, the splitting is significant enough that the resulting line profile is poorly reproduced with a single Gaussian profile. At the typical magnetic field strengths expected in the outer regions of protoplanetary disks, on the order of 1 -- 10~mG, the Zeeman split lines still maintain a close-to-Gaussian morphology, as discussed more in Appendix~\ref{sec:app:model:zeeman}.

Figure~\ref{fig:zeeman_splitting_demonstration} also highlights how the projection of the magnetic field along the line of sight changes the resulting profile. When aligned with the line of sight, $\theta = 0\degr$, only the $\sigma^{\pm}$ transitions populate. With a progressively more perpendicular field, $\theta$ approaches $90\degr$, the $\pi^0$ transitions starts to dominate. As only the $\sigma^{\pm}$ transition groups are shifted relative to the true line center, broadening is more significant when the magnetic field is preferentially aligned with the line of sight.

The final component of the model was a magnetic field prescription. As the magnitude of the Zeeman splitting is sensitive to $1 + \cos^2(\theta)$, where $\theta$ is the angle between the magnetic field vector and the line of sight \citep{Lankhaar_Teague_2023}, we explore two scenarios: one where $\theta$ is constant around the annulus, as would be the case for some large-scale, external field, and one where $\theta$ is able to vary as a function of $\phi$ in a manner which can be recast as an azimuthally symmetric disk-centric field described by $(B_{\phi},\, B_r,\, B_z)$.

These two scenarios are described by $|\mathbf{B}|$, the magnetic field strength, $\theta_B$, the azimuthally averaged $\theta$ value, and, for the disk-centric case, $\phi_B$, the mixing angle between toroidal and radial magnetic fields. For the latter case, these can be recast to describe the disk-centric components,
\begin{align}
    B_{\phi} &= |\mathbf{B}| \sin(\theta_B) \cos (\phi_B),\\
    B_r      &= |\mathbf{B}| \sin(\theta_B) \sin (\phi_B),\\
    B_z      &= |\mathbf{B}| \cos(\theta_B),
\end{align}
where all values are implicitly dependent on radius and from which $\theta$ can be calculated for any position around the annulus:
\begin{equation}
    |\mathbf{B}|\cos\big(\theta(\phi)\big) = B_{\phi} \cos(\phi) \sin(|i|) - B_{r} \sin(\phi) \sin(i) - B_z \cos(i) .
\end{equation}
This seemingly more complex parameterization was chosen over modeling $(B_{\phi},\, B_r,\, B_z)$ directly, as uniform priors in these parameters results in non-uniform priors for $|\mathbf{B}|$ and $\theta_B$, that, most importantly, favor non-zero magnetic field strengths. The additional benefit is that the inferred posteriors for $|\mathbf{B}|$ and $\theta_B$ can be directly compared between the two field morphologies.


In total, the model contains either nine or ten free parameters: three describing the azimuthally symmetric velocity structure of the disk, $v_{\phi}$, $v_r$, and $v_z$; four describing the excitation conditions of the line, $T_{\rm kin}$, $N({\rm CN})$, $n({\rm H_2})$, and $\Delta V_{\rm nt}$; either two or three describing the magnetic field, $(|\mathbf{B}|,\, \theta_B,\, \phi_B)$, and one describing the width of the convolution kernel, ${\rm FWHM_{conv}}$.

With this model, we implicitly ignore the impact of continuum opacity, motivated by two reasons. Firstly, with an angular resolution of ${\approx}~0\farcs45$, the observations used here are minimally affected by convolution effects beyond ${\approx}~0\farcs9$ and so the analysis will necessarily focus on those regions of the disk. At these radii, \citet{Macias_ea_2021} estimates the 3.1~mm continuum optical depth to be $\lesssim 0.1$, suggesting that effects from the continuum will be negligible. Secondly, \citet{Lankhaar_Teague_2023} argued that while continuum opacity would suppress polarized emission, it would only marginally impact the broadening of unpolarized emission. This is because the continuum subtraction effect is wavelength independent, at least across the bandwidths needed for excitation analyses, and so would not result in a changing of the line width ratios.

\subsection{MCMC Sampling}
\label{sec:methodology:mcmc}

An exploration of the parameter posterior distributions was performed with the MCMC package \texttt{emcee} \citep{emcee} utilizing 128 walkers which took 20,000 steps using the default \texttt{StretchMove} move. The median autocorrelation time of all parameters, $\langle \tau \rangle$, was estimated to be ${\sim}200$. As such, $20\langle \tau \rangle$ samples were removed as burn-in steps with the remaining samples representing the posterior distributions. To create the radial profiles, annuli were extracted every $0\farcs1$, roughly a fifth of the beam major axis, with a width of one pixel, $0\farcs05$, between $0\farcs4$ and $4\arcsec$ from the star. The narrow annuli are necessary to reduce the impact of Keplerian shear across the annulus width that would introduce significant non-thermal broadening at a level that ranges between ${\sim}~100~{\rm m\,s^{-1}}$ at $1\arcsec$ down to ${\sim}~10~{\rm m\,s^{-1}}$ at 3\arcsec{} \citep[e.g.,][]{Teague_Loomis_2020}. All parameters assumed uniform priors over some broad range of possible values, except the non-thermal broadening contribution which was chosen to have a Gaussian prior centered on zero with a width of $0.08~c_s$ based on the upper limits reported in \citet{Flaherty_ea_2018}.

\subsection{Randomized Zeeman Sensitivities}
\label{sec:methodology:broadening}

\begin{figure*}
    \includegraphics[width=\textwidth]{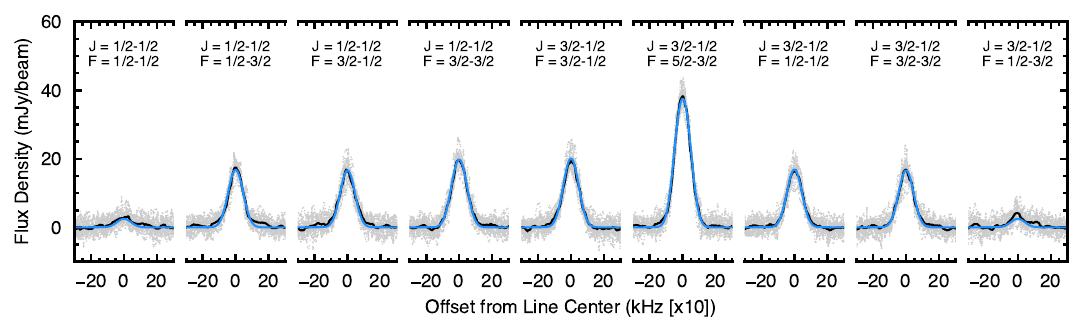}
    \caption{Example of the velocity-deprojected spectrum at $1\farcs4$ (gray points with the black line showing a running median with a kernel width of $80~{\rm m\,s^{-1}}$) and the best-fit model including a disk-centric magnetic field morphology in blue. The model is able to reproduce the data with an exceptional fidelity.}
    \label{fig:model_spectrum}
\end{figure*}

Zeeman splitting offers a powerful way to differentiate the broadening due to magnetic fields from other broadening processes, as only certain lines are sensitive to the Zeeman effect. In the model described above, the thermal and non-thermal broadening, as  well as the smoothing convolution, will uniformly broaden all hyperfine components. In contrast, increasing the magnetic field strength will broaden certain transitions more than others. To verify that the broadening we detect can be uniquely associated with magnetic-field induced Zeeman splitting, we run two additional `randomized' models for both field morphology scenarios where the Zeeman-specific properties of the most sensitive transitions, $J=3/2 - 1/2$ $F = 3/2 - 1/2$ and $J = 1/2 - 1/2$ $F = 3/2 - 3/2$, are swapped with those of the least sensitive transitions $J=3/2 - 1/2$ $F = 1/2 - 1/2$ and $J=3/2 - 1/2$ $F = 3/2 - 3/2$. The two models represent the two possible pairs of swapping. In this case, the inferred magnetic field should prefer lower values as to not over-broaden the less sensitive, and thus narrower, transitions.

As this approach does not completely remove all sensitivity to magnetic fields, it is not expected that a field strength consistent with zero is recovered. Instead, we should infer a strength that is significantly lower than with the correct molecular properties. To classify the posterior distributions as statistically different, we require that their medians are separated by more than the quadrature sum of their standard deviations. Appendix~\ref{sec:app:model} shows the posterior distributions for each radius in each case.

\section{Results}
\label{sec:results}

In this section, we discuss the inferred magnetic field strength, morphology, and limits on the plasma beta. A more comprehensive discussion of the derived velocity profiles and excitations conditions can be found in Appendix~\ref{sec:app:model}. Figure~\ref{fig:model_spectrum} demonstrates the typical quality of fit we are able to achieve with the model. In general, beam smearing will impact the shape of the spectrum in regions inwards of approximately twice the beam FWHM. This effect heavily biases the results interior to ${\approx}~1\arcsec$ (60~au), so we are cautious over interpretation of the posteriors in this region.

\subsection{Field Strength}
\label{sec:results:strength}

\begin{figure*}
    \centering
    \includegraphics[width=\textwidth]{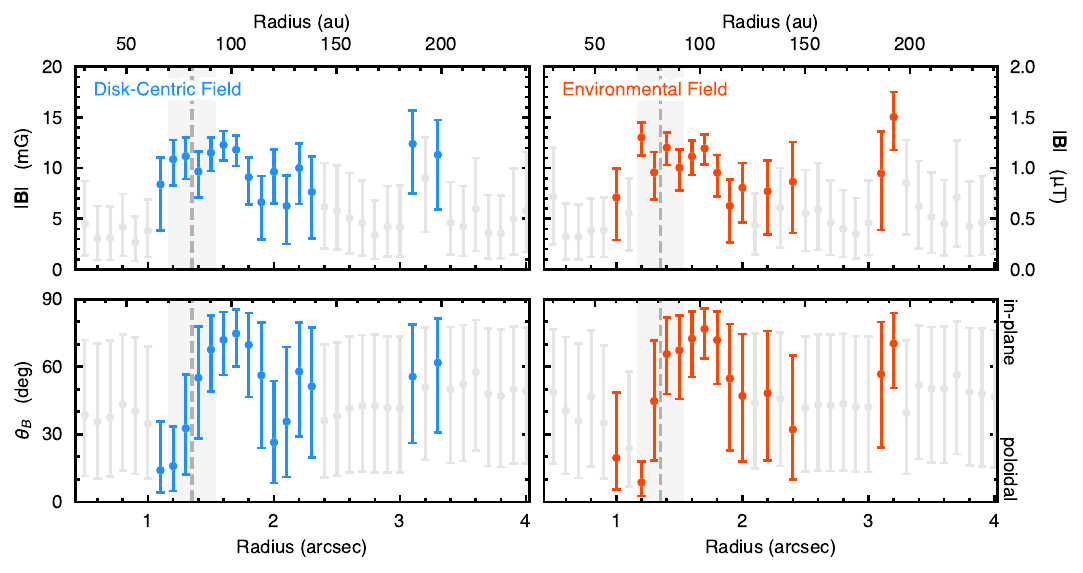}
    \caption{The inferred magnetic field strength, $|\mathbf{B}|$ top row, and the azimuthally averaged angle between the magnetic field and the line of sight, $\theta_B$, bottom row. The left column shows the results when adopting a disk-centric description of the magnetic field where $\theta$ varies as a function of $\phi$, while the right column shows the same when assuming an environmental field where $\theta$ is constant. The error bars represent the 16th or 84th percentile range about the posterior median. Gray points show radii where Zeeman-broadening cannot be robustly identified as the source of broadening, as discussed in Section~\ref{sec:methodology:broadening}. The vertical dashed line and gray shaded region show the location of the outer continuum gap reported by \citet{Ilee_ea_2022} and \citet{Das_ea_2024}.}
    \label{fig:B_theta_radial_profiles}
\end{figure*}

The inferred magnetic field strengths and (azimuthally averaged) field orientations are shown in Figure~\ref{fig:B_theta_radial_profiles} for the two magnetic field parameterizations. The colored points show radii where the magnetic field strength posteriors of the `correct' Zeeman sensitivities are significantly higher than for the `mixed' Zeeman sensitives. For completeness, we plot posterior distributions for all radii where we were able to obtain a constraint on the gas temperature and CN column density.

Both field morphologies yield a similar radially averaged magnetic field strength of $9 \pm 2$~mG and $8 \pm 2$~mG for the disk-centric model and for the environmental field, respectively. Both models also recover a comparable radial profile for the magnetic field strength: a negligible field within the inner $1\arcsec$ (60~au) and beyond ${\sim}~2\farcs5$ (150~au), and a highly significant, $|\mathbf{B}| \, / \,\delta|\mathbf{B}| \gtrsim 5$, where $\delta|\mathbf{B}|$ is the uncertainty on $|\mathbf{B}|$, detection of a ${\sim}~10~{\rm mG}$ ($1~\mu{\rm T}$) field between $1$ and $2\farcs5$. For the disk-centric case, a magnetic field is also detected at $3\farcs2$ (190~au) with a ${\sim}~12$~mG field strength, however such a field is not found for the environmental scenario. Interestingly, the magnetic field strength appears to peak at the location of a gap observed in scattered light, thermal continuum, and molecular line emission at $1\farcs37$ \citep[80~au;][]{vanBoekel_ea_2017, Ilee_ea_2022, Teague_ea_2017}, which will be discussed further in Section~\ref{sec:discussion:field_morphology}.

\subsection{Field Orientation}
\label{sec:results:orientation}

\begin{figure*}
    \centering
    \includegraphics[width=\textwidth]{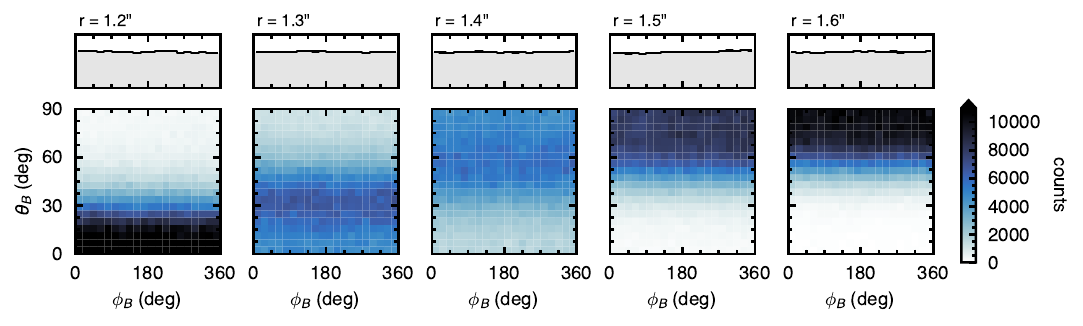}
    \caption{The posterior distributions of $\theta_B$ and $\phi_B$ across the radial region where the magnetic field was robustly detected.  Is is clear that the field transitions from predominantly poloidal ($\theta_B = 0\degr$) at $1\farcs1$ to one in the plane of the disk ($\theta_B = 90\degr$) by $1\farcs5$. Across this radial range there is no clear preference for a $\phi_B$ value with the posterior distribution equally sampling the allowed range as shown in the histograms in the top row.}
    \label{fig:B_covariances}
\end{figure*}

The inferred field morphologies are also consistent between the two assumed morphologies, in so far as the data has little constraining power for this parameter. The bottom row of Fig.~\ref{fig:B_theta_radial_profiles} shows the posterior distributions for $\theta_B$, the azimuthally averaged value of $\theta$. Note that for the disk-centric model, $\theta$ varies as a function of annulus, while for the environmental model, $\theta$ is constant as a function of azimuth. In both cases, a predominantly poloidal field is inferred interior to the gap center at $1\farcs37$, which rapidly transitions to one within the plane of the disk at larger radii. Beyond $1\farcs8$ (110~au), the data is insufficient to constrain $\theta_B$ (with the posterior distributions matching the uniform priors), except at $3\farcs2$, where the disk-centric model recovers a magnetic field and an in-plane field is inferred.

In the disk-centric model, $\phi_B$, the mixing angle between the $B_{\phi}$ and $B_r$ components, was an additional free parameter. At no radius in the disk was a constraint on this parameter inferred. The modeled posterior distributions are entirely consistent the uniform prior distribution which spanned $0\degr$ to $360\degr$. Figure~\ref{fig:B_covariances} demonstrates this by plotting the covariances between $\theta_B$ and $\phi_B$ for five radial bins where the magnetic field was robustly detected. It is clear that there is a steady transition from $\theta_B \lesssim 30\degr$ at $1\farcs2$ to $\theta_B \gtrsim 60\degr$ at $1\farcs6$, while at each radius $\phi_B$ is completely unconstrained. This suggests that either the data sensitivity and/or spectral resolution is insufficient to tease out this subtle variation in broadening or that there is no variation in $\theta$ as a function of $\phi$. The interpretation of this in terms of a global field morphology is discussed in Section~\ref{sec:discussion:field_morphology}.

\subsection{Plasma $\beta$}
\label{sec:results:plasmab}

The excitation analysis also provides an opportunity to derive a radially resolved plasma $\beta$ value, the ratio of the thermal to magnetic pressure, from where the CN emission arises. The thermal pressure was calculated assuming an ideal gas with a temperature set to the inferred $T_{\rm kin}$ and a gas density given by the $n({\rm H_2})$ collider density. As discussed in Appendix~\ref{sec:app:model}, the data favored an LTE solution at all radii, which results only in lower limits of the gas density and, in turn, the derived plasma $\beta$. The magnetic pressure is given by $|\mathbf{B}|^2 \, / \, 8\pi$ \citep{Lesur_ea_2023}.

\begin{figure}
    \includegraphics[]{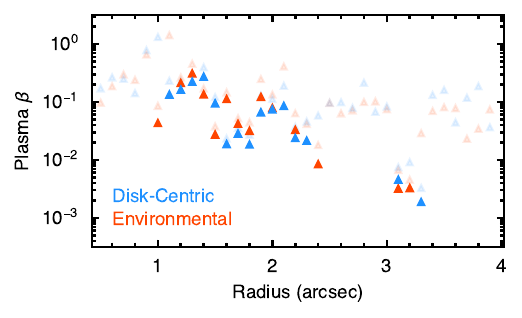}
    \caption{Lower limits to the plasma $\beta$ in the vertical region from where the CN emission emanates, where the median of the $\beta$ posteriors were plotted. The blue and red points represent the disk-centric and environmental field morphologies, respectively, while the more transparent points are where Zeeman broadening cannot be robustly inferred.}
    \label{fig:plasma_beta}
\end{figure}

The resulting $\beta(r)$ radial profiles are shown in Fig.~\ref{fig:plasma_beta} where, the median of the parameter distribution was chosen as the plotted lower limit, and semi-transparent points mark where the inferred broadening cannot be robustly attributed to Zeeman splitting. Both field morphologies yield similar limits, with $\beta$ ranging between ${\gtrsim}0.1$ in the inner disk and values as low as ${\sim}0.01$ at the disk edge, albeit with increased scatter. It is important to note that these values are \emph{not} for the midplane, but rather from the CN emission layer. Based on the gas kinetic temperature, discussed in Appendix~\ref{sec:app:model}, emission interior to $2\farcs5$ (150~au) traces regions that are warmer than 10~K (up to 30~K in the inner disk regions), so likely substantially above the midplane. Indeed, for the case of IM~Lup, \citet{Paneque-Carreno_ea_2024} report CN $N=3-2$ emission arising from a region that is ${\sim}~0.2r$ above the midplane. Beyond this region, the recovered kinetic temperature plateaus at 10~K, potentially signaling a region closer to the disk midplane; however, given the quality of the signal at these extended radii, this should be taken with caution.

\section{Discussion}
\label{sec:discussion}

\subsection{Global Magnetic Field Morphology}
\label{sec:discussion:field_morphology}

Two different field morphologies were invoked to explain the observations: a disk-centric field which could be decomposed into azimuthally symmetric components $(B_{\phi},\, B_r,\, B_z)$, such that $\theta$ varied across the surface of the disk; and a large-scale environmental field, where $\theta$ was assumed to be constant as a function of azimuth. As both morphologies were able to equally well explain the data, and recovered similar field strengths and (azimuthally averaged) angles between the line of sight and the magnetic field vector, neither is clearly favored.

For the disk-centric morphology, the model was flexible enough to allow toroidal and radial magnetic field components (with most models predicting strong toroidal components close to the disk midplane; \citealt{Flock_ea_2015}). However, as demonstrated in Figure~\ref{fig:B_covariances}, no distinction was possible. This could be due to two different issues. Firstly, the data could be too noisy to identify the subtle azimuthal variations in broadening that would allow the distinction between $B_{\phi}$ and $B_r$ components. The azimuthal variations in $\theta$ that arise due to the changing projection of $B_{\phi}$ and $B_r$ components are only ${\sim}10\degr$, much lower than the typical width of the posterior on $\theta_B$, around ${\sim}30\degr$, suggesting that this is likely. Note that in more moderately inclined sources, this variation will be larger, in principle making the distinction easier, however in practice the added complications of contamination from the bottom surfaces will be the limiting factor \citep[e.g.,][]{Izquierdo_ea_2025}.

A second possibility is that we are instead tracing a large-scale environmental field, which should have an azimuthally constant $\theta$ that results in a azimuthally varying $\phi_B$. In this scenario, it would be expected that, as the annuli are symmetric about the star, both the field strength and $\theta$ are constant as a function of radius (the magnetic field should not know about the orientation and position of the disk). While the magnetic field strength could plausibly be argued to be constant as a function of radius given the large error bars, there is a clear variation in the morphology of the field. A scenario in which the gapped disk structure perturbs the environmental field and alters its morphology \citep[just as disks have been shown to for local field, e.g.,][]{Riols_Lesur_2019} could account for this, however deeper observations that are able to constrain the magnetic field strength and orientation across a larger radial range would be necessary to truly determine the global field morphology.

\subsection{Comparisons with Previous Studies}
\label{sec:discussion:previous_studies}

There have been many attempts to detect magnetic fields in protoplanetary disks.  These can broadly be split between those searching for polarized continuum emission arising due to magnetic grain alignment, and those attempting to detect linear polarized molecular emission from with the Goldreich Kylafis effect \citep[GK;][]{Goldreich_Kylafis_1982} or the circular polarized emission from Zeeman splitting.

Thus far, there has only been one claimed detection of a magnetic field in a protoplanetary disk. \citet{Ohashi_ea_2025} report the presence of a primarily toroidal magnetic field threading the disk around HD~142527. Based on the coherence of the morphology of the polarization, the authors argued that magnetic grain alignment was the dominant polarization mechanism and recovered a field strength of ${\sim}0.3$~mG at a distance of ${\sim}200$~au when comparing the inferred magnetic field morphology to specific MHD predictions. This field is considerably weaker than the field reported here, by almost two orders-of-magnitude. When assuming a local gas density of $6.6 \times 10^{-16}~{\rm g\,cm^{-3}}$, around the same value as the lower limit we recover for TW~Hya, the authors find a plasma $\beta$ of ${\sim}200$ resulting in $\alpha$ values of ${\sim}10^{-3}$, broadly consistent with typical levels of turbulence found via non-thermal broadening \citep[e.g.,][]{Flaherty_ea_2020} or the vertical extent of dust rings \citep{Doi_ea_2023}. It should be noted, however, that the $\beta$ derived here would be representative of the vertical region from which CN emanates, while that reported in \citet{Ohashi_ea_2025} will be describing the midplane. As described in Appendix~\ref{sec:app:model:excitation}, the excitation conditions are suggestive of a vertical region between $z \, / \, r \sim 0.2$ -- 0.3, consistent with the findings of \citet{Teague_Loomis_2020} and \citet{Yoshida_ea_2024}.

HD~142527 also exhibited some level of polarized CO emission after \citet{Stephens_ea_2020} followed a `shift-and-stack' process to uncover a marginal detection. Intriguingly, their Figure~A5 shows that $^{13}$CO shows a strong polarized signal peaking at the location where \citet{Ohashi_ea_2025} found evidence of magnetic grain alignment. Unfortunately the GK effect is not sensitive to the magnetic field strength, and any sensitivity to the field morphology, which would need extensive numerical modeling to interpret, was removed due to the shift-and-stack analysis. Nonetheless, the presence of linearly polarized molecular emission coincident with regions where magnetic grain alignment is found to be the dominant polarization mechanism strongly suggest that even marginal detections of polarized molecular emission can signal a magnetic field.

For the case of TW~Hya, \citet{Vlemmings_ea_2019} placed tight upper limits on the level of circularly polarized emission from CN. Despite a robust detection of the unpolarized emission, no polarized emission was found: the authors converted those results to a $1\sigma$ upper limit on the magnetic field strength in the disk of $|B_z| \leq 0.8~{\rm mG}$ or $B_{\phi} \leq 30~{\rm mG}$. While the toroidal limits are consistent with the results presented here, the poloidal field limits are over an order-of-magnitude weaker than what is inferred around $1\farcs2$ based on the Zeeman broadening (where $|B_z| \sim 11~{\rm mG}$). It is impossible to definitively determine why circular polarization was not detected, however it is likely that cancellation of the polarized signal, and not the broadening signal, is the most likely culprit. Cancellation can be intrinsic to the source or due to the chosen observational setup. If there is a change in the magnetic field morphology along the path traced by the emission, then alternating field orientations can result in cancellation. This can happen at the very local level, with a highly tangled or turbulent magnetic field \citep[e.g.,][]{Hughes_ea_2009}, or due to large, disk-wide variations, such as the flip in direction either side of the disk midplane if a toroidal field dominates \citep{Lankhaar_Teague_2023}. If polarized emission does escape from the disk, then it will exhibit a highly structured morphology \citep{Mazzei_ea_2020}. Without sufficient angular resolution to spatially resolve these substructures, the signal will be averaged over and lie undetected. If a circular polarization is able to be detected, potentially with low frequency, high angular resolution observations to remove the continuum absorption and spatial cancellation, then the relative importance of these effects can be constrained.

However, linearly polarized molecular emission has been reported for TW~Hya: \citet{Teague_ea_2021} found low-level linearly polarized emission from $^{12}$CO and $^{13}$CO between $0\farcs5$ and $1\arcsec$ which would require the presence of a (very weak) magnetic field in order for the GK effect to work. As for the case of HD~142527, the shift-and-stack methodology removes any information about the magnetic field structure. As these works have shown, prior assumptions about the field morphology are needed in order to tease out the weak polarized signals. A re-analysis of these datasets with a prior model of the underlying magnetic field based on the results presented here would offer an opportunity to test our understanding of polarized radiative transfer and whether the limits based on the lack of circular emission should be reevaluated.

\subsection{Comparison with the Young Solar System}
\label{sec:discussion:solar_system}

The inferred magnetic field strengths presented here can be compared to those from paleomagnetic measurements of meteorites from the early solar system and to expectations from MHD simulations of protoplanetary disks. Most paleomagnetically studied meteorites likely originate from the innermost ${\sim}10$~au of the solar system \citep{Sutton_ea_2017} and none are thought to sample distances ${\gtrsim}50$~au, precluding direct comparison with the results here. However, meteorite-inferred magnetic fields are relevant in three ways. First, magnetic field strengths between 10 and 1000~mG recorded by meteorites likely sampling the 2--7 au region broadly support the importance of magnetic field-driven radial transport, which predicts magnetic field of order 10--100~mG in this region \citep{Wardle_2007, Bai_2015}. Second, these models, which are derived by relating the magnetic torque to the observed inward disk accretion rate, would predict magnetic fields of order 1~mG at 60 au, assuming a $2.5 \times 10 ^{-9}$ $M_\odot \, {\rm yr}^{-1}$ accretion rate \citep{Herczeg_ea_2023}. These predictions are supported by inferences of weak fields, typically upper limits of ${\lesssim}0.1~\mathrm{G}$, from paleomagnetic studies of the most distantly sourced (${\gtrsim}7~\mathrm{au}$) material \citep[e.g.,][]{Mansbach_ea_2025}. The ${\sim}10$~mG field inferred here is ${\sim}10\,{\times}$ stronger than those predicted by these models, which may indicate local magnetic field amplification processes, such as spontaneous zonal field concentration \citep{Suriano_ea_2018}, unrelated to large-scale radial transport. 

This leads to the third inference from solar system meteorites, which have provided evidence for the persistence of strong magnetic field ($\geq 1000$~mG) zones in the otherwise weakly magnetized carbonaceous chondrite-formation regions \citep{Fu_ea_2021, Borlina_ea_2021}. If hypothesized magnetic field enhancement mechanisms \citep[e.g.,][]{Suriano_ea_2018, Riols_Lesur_2019} can result in the ${\sim}100{\times}$ increase in local magnetic fields in the solar system, they would be sufficient for producing 10~mG fields observed in TW Hya, assuming they can operate at such large radii and low density, high-ionization conditions. 

\subsection{Future Outlook}

The implementation of the Wideband Sensitivity Upgrade \citep[WSU;][]{Carpenter_ea_2023} for ALMA will facilitate extremely high spectral resolution observations with the default spectral resolution dropping to 13.5~kHz ($36~{\rm m\,s^{-1}}$ at the 113~GHz of the ground-state CN transition), and `zoom' modes offering even finer resolutions. This improvement will bring huge power through the enhanced sampling of the narrow CN emission and enable even more subtle variations in line width to be discerned. Furthermore, the 2~GHz bandwidth available at this high spectral resolution will provide an opportunity to target multiple Zeeman-sensitive species simultaneously, vastly improving the information content of these data. 

While new instrumentation will increase our sensitivity to weaker magnetic fields, a more widespread issue is the contamination of the spectra from emission from the back side of the protoplanetary disk for all sources that are not viewed face on. This has already been acknowledged as an issue when looking at the kinematics of moderately inclined sources \citep[e.g.,][]{Izquierdo_ea_2025}. While not insurmountable, accounting for the 3D nature of the emitting source will require the use of extensive radiative transfer models and the development of techniques able to segment the data into contributions from the top and bottom surfaces.

\section{Conclusions}
\label{sec:conclusions}

We have presented a novel approach to detecting and characterizing magnetic fields through the subtle broadening of molecular emission lines due to Zeeman splitting. Application to observations of CN $N=1-0$ emission from the disk of TW~Hya reveal the presence of a ${\sim}10$~mG field between $1\arcsec$ (60~au) and $2\arcsec$ (120~au), apparently peaking in strength at the radial location of a well-characterized gap in the disk at ${\sim}80$~au. The data were insufficient to provide tight constraints on the morphology of the field, however the modeling favors a predominantly poloidal field interior to the gap and one that is within the plane of the disk exterior to it. The data were unable to support a decomposition of the in-plane contribution into toroidal and radial components. However, the data were equally well fit with a model that imposed an azimuthally constant $\theta$ highly suggestive of the magnetic environment at these radii being dominated by a large-scale environmental field rather than one generated by dynamos within the star and/or disk. An excitation analysis, yielding lower limits to the gas pressure, and the derived magnetic field strength constraints allowed for strict lower limits on a radially resolved plasma $\beta$ value ranging between $10^{-1}$ and $10^{-2}$. This approach to detecting magnetic fields without the need for polarized emission offers a powerful angle with which to attack the question of the  importance of magnetic fields for the formation of planetary systems.

\begin{acknowledgments}
We thank the referee for helpful comments which significantly improved the clarity of the manuscript. This paper makes use of the following ALMA data: ADS/JAO.ALMA\#2022.1.00840.S. ALMA is a partnership of ESO (representing its member states), NSF (USA) and NINS (Japan), together with NRC (Canada), MOST and ASIAA (Taiwan), and KASI (Republic of Korea), in cooperation with the Republic of Chile. The Joint ALMA Observatory is operated by ESO, AUI/NRAO and NAOJ. The National Radio Astronomy Observatory and Green Bank Observatory are facilities of the U.S. National Science Foundation operated under cooperative agreement by Associated Universities, Inc. BL acknowledges support from the Swedish Research Council (VR) under grant number 2021-00339
\end{acknowledgments}

\vspace{5mm}
\facilities{ALMA}

\software{\texttt{CASA} \citep{CASA}, \texttt{RADEX} \citep{vanderTak_ea_2007}, \texttt{SpectralRadex} \citep{SpectralRadex}, \texttt{GoFish} \citep{GoFish}}

\begin{appendix}

\section{Molecular Properties}
\label{sec:app:molecular_properties}

\begin{deluxetable*}{cccccccccc}
\tablecaption{CN $N=1-0$ Hyperfine Components \label{tab:observations}}
\tabletypesize{\footnotesize}
\tablehead{
\colhead{$J^{\prime} - J^{\prime\prime}$} &
\colhead{$F^{\prime} - F^{\prime\prime}$} &
\colhead{$\nu_0$} &
\colhead{$A_{\rm ul}$} &
\colhead{$g_{\rm u}$} &
\colhead{$z$} &
\colhead{$\Delta Q$} &
\colhead{$\bar{Q}$} &
\colhead{$w$} &
\colhead{Integrated Intensity}\\
\colhead{} &
\colhead{} &
\colhead{$({\rm GHz})$} &
\colhead{$({\rm s^{-1}})$} &
\colhead{} &
\colhead{$({\rm kHz \, mG^{-1}})$} &
\colhead{} &
\colhead{} &
\colhead{} &
\colhead{$({\rm Jy\,km\,s^{-1}})$}
}
\decimals
\startdata
1/2 - 1/2 & 1/2 - 1/2 & 113.123369 & $1.29 \times 10^{-6}$ & 2 & -0.62           & \phantom{-0}3.01 & \phantom{0}5.01 & 0.038 & $0.101 \pm 0.009$ \\
1/2 - 1/2 & 1/2 - 3/2 & 113.144147 & $1.05 \times 10^{-5}$ & 2 & \phantom{-}2.18 & \phantom{0}-0.98 & \phantom{0}1.14 & 0.030 & $0.476 \pm 0.009$ \\
1/2 - 1/2 & 3/2 - 3/2 & 113.191289 & $6.68 \times 10^{-6}$ & 4 & \phantom{-}0.63 & \phantom{-}21.89 & 42.84           & 0.326 & $0.565 \pm 0.009$ \\
1/2 - 1/2 & 3/2 - 1/2 & 113.170511 & $5.14 \times 10^{-6}$ & 4 & -0.30           & \phantom{-0}0.06 & \phantom{0}8.42 & 0.031 & $0.462 \pm 0.009$ \\
3/2 - 1/2 & 3/2 - 1/2 & 113.488126 & $6.74 \times 10^{-6}$ & 4 & \phantom{-}2.17 & \phantom{-}15.07 & 17.07           & 0.447 & $0.569 \pm 0.009$ \\
3/2 - 1/2 & 5/2 - 3/2 & 113.490964 & $1.19 \times 10^{-5}$ & 6 & \phantom{-}0.56 & \phantom{0}-0.96 & \phantom{0}1.28 & 0.009 & $1.148 \pm 0.009$ \\
3/2 - 1/2 & 1/2 - 1/2 & 113.499634 & $1.06 \times 10^{-5}$ & 2 & \phantom{-}0.62 & \phantom{0}-0.86 & \phantom{0}1.97 & 0.015 & $0.486 \pm 0.009$ \\
3/2 - 1/2 & 3/2 - 3/2 & 113.508902 & $5.19 \times 10^{-6}$ & 4 & \phantom{-}1.62 & \phantom{0}-0.86 & \phantom{0}1.26 & 0.025 & $0.455 \pm 0.009$ \\
3/2 - 1/2 & 1/2 - 3/2 & 113.520420 & $1.30 \times 10^{-6}$ & 2 & \phantom{-}1.56 & \phantom{0}-0.53 & \phantom{0}4.27 & 0.080 & $0.102 \pm 0.009$ \\
\enddata
\end{deluxetable*}

Table~\ref{tab:observations} details the observed molecular transitions, pertinent Zeeman-related properties and their observed integrated intensities. Einstein A coefficients, $A_{\rm ul}$, and upper-state level weights, $g_{\rm u}$, were taken from CDMS \citep{Endres_ea_2016} which collates data from \citet{Dixon_Woods_1977} and \citet{Skatrud_ea_1983}. Properties relating to Zeeman splitting, $z$, $\Delta Q$ and $\bar{Q}$, were taken from \citet{Lankhaar_Teague_2023}.

\begin{figure*}
    \centering
    \includegraphics[width=\textwidth]{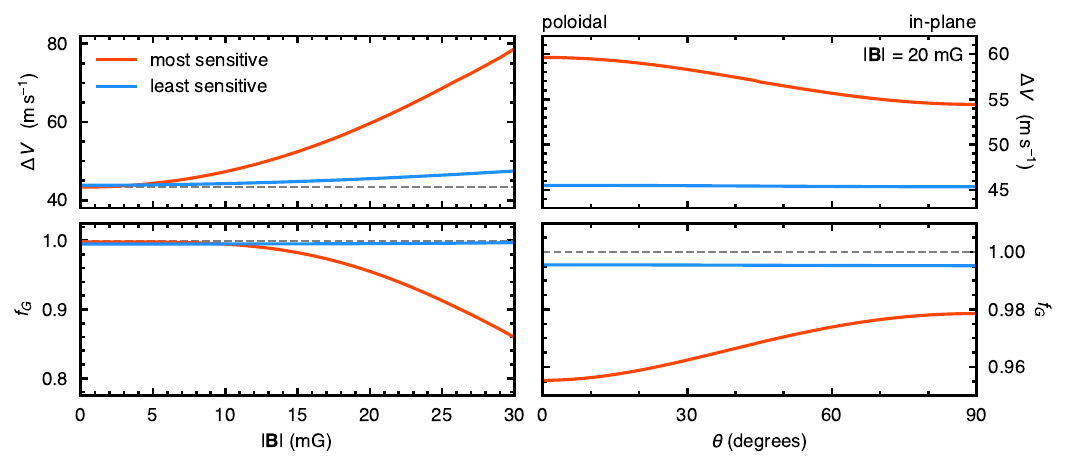}
    \caption{Demonstrating the different responses to changing the magnetic field strength, left column, and orientation, right column, the most sensitive hyperfine component, red, and the least sensitive hyperfine component, blue, exhibit. The top row shows the change in line width, as measured by the fit of a Gaussian line profile. The bottom row shows the deviation from a Gaussian line profile, quantified by $f_G$ which describes the fraction of the total intensity which can be described by a Gaussian profile. The other seven hyperfine components will have profiles which fall between these two limiting cases.}
    \label{fig:app:broadening_nongaussianity}
\end{figure*}

The line centers were measured from the data. Firstly, spectra from each annulus was aligned and averaged using the velocity profiles described in Appendix~\ref{sec:app:model} before having a Gaussian line profile fit to it to derive a line center. The mean line center between $0\farcs5$ and $3\arcsec$ was taken as the true line center after subtracting a systemic velocity of $2.84~{\rm km\,s^{-1}}$. The typical uncertainty on each of these is less than 10~kHz and is dominated by the systematic uncertainty related to the determination of the systemtic velocity.

The integrated intensities were calculated using standard `shift-and-stack' techniques \citep[e.g.,][]{Yen_ea_2016} where a dynamical stellar mass of $0.84~M_{\odot}$ and a disk inclination of $5\fdg 8$ were adopted. The data were spatially integrated out to $4\arcsec$ (240~au) and spectrally over $\pm 300~{\rm kHz}$, or equivalently $\pm \, {\sim}800~{\rm m\,s^{-1}}$ at this frequency.

\section{Observational Signatures of Zeeman Broadening}
\label{sec:app:zeeman_broadening}

To robustly identify the presence of a magnetic field through a change in the line profile caused by Zeeman splitting without a companion polarization measurement, we must be confident that the broadening we see can be uniquely attributed to Zeeman splitting. In this Appendix, we discuss how this broadening manifests for the low to moderate fields expected in protoplanetary disks, and how we can be confident that we are correctly identifying a magnetic field.

As described in Section~\ref{sec:methodology:zeeman} and shown in Figure~\ref{fig:zeeman_splitting_demonstration}, the broadening experienced by each line results from the separation of the three magnetic subtransition groups, $\pi^0$, $\sigma^+$ and $\sigma^-$, dependent on the magnetic field strength, $|\mathbf{B}|$, and the relative propagation propensity of these transition groups, influenced by the magnetic field orientation, $\theta$. The sensitivity of each line to these perturbations, initially derived in \citet{Crutcher_ea_1996}, is codified in the line-specific parameters $z$, $\Delta Q$, and $\bar{Q}$, listed in Table~\ref{tab:observations} and described in \citet{Lankhaar_Teague_2023}. At the moderately low field strengths expected to be threading protoplanetary disks, the dominant signal will therefore be a specific ratio of line widths; more sensitive lines should experience more broadening than less sensitive lines.

This is demonstrated in the top row of Figure~\ref{fig:app:broadening_nongaussianity} which shows the change in line width (derived by the fit of a Gaussian profile) when the magnetic field is increased for the most sensitive hyperfine component to Zeeman splitting, $N=1-0$ $J=3/1-1/2$ $F=3/2-1/2$, in red, and the least sensitive hyperfine component, $N=1-0$ $J=3/1-1/2$ $F=5/2-3/2$, in blue. The left hand panel assumes a purely poloidal field $\theta = 0\degr$ and varies $|\mathbf{B}|$, while the right panel fixed $|\mathbf{B}| = 20~{\rm mG}$ and varies $\theta$. Note that the remaining 7 hyperfine components will fall between these two limiting cases. This demonstrates that each magnetic field configuration, $(|\mathbf{B}|,\, \theta)$, should relate to a very specific ratio of the line widths.

With stronger fields, the Zeeman splitting can become sufficiently strong that the lines deviate from a purely Gaussian profile. This impact of this is shown in the bottom row of Figure~\ref{fig:app:broadening_nongaussianity} which plots $f_G$, a parameter which describes the non-Gaussianity of the line. This parameter is defined as $f_G = 1 - (\sum | {\rm I_{\rm instrinic}} - I_{\rm gaussian}| \, / \sum I_{\rm intrinsic})$, where $I_{\rm intrinsic}$ is the intrinsic line profile and $I_{\rm gaussian}$ is the best-fit Gaussian profile. When a Gaussian profile perfectly reproduces the intrinsic line profile, $f_G = 1$, and drops as the line becomes progressively less Gaussian. As can be seen, it is only at magnetic field strengths around 15~mG that a noticeable deviation from Gaussianity is observed (although the specific field strength this occurs at will depend on the optical depth of the line as the saturate core will also lead to non-Gaussianity). Importantly, the methodology described in Section~\ref{sec:methodology} models the full Zeeman splitting and so implicitly accounts for any non-Gaussianity in the line, either through the Zeeman splitting or from optical depth effects.

\begin{figure}
    \centering
    \includegraphics[width=\linewidth]{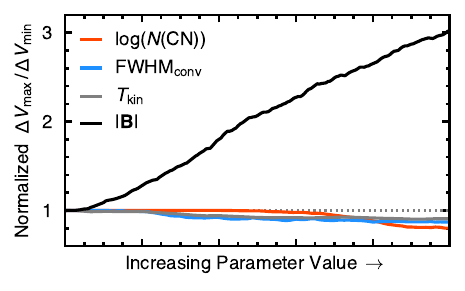}
    \caption{Showing how the ratio of line widths between the most sensitive hyperfine component and the least sensitive hyperfine component changes as various parameters of the model are increased. Only when the magnetic field strength, $|\mathbf{B}|$, is increased, does the ratio increase; for all other parameters the ratio decreases. This suggests that the signature of Zeeman splitting cannot be mimicked by other model properties. The parameter ranges for the four properties are 0 - 50~mG for $|\mathbf{B}|$, 20 - 50~K for $T_{\rm kin}$, 0 - $80~{\rm m\,s^{-1}}$ for ${\rm FWHM_{conv}}$ and 11 - 15 for $\log(N({\rm CN}))$.}
    \label{fig:app:broadening_sensitivity}
\end{figure}

Another fortuitous aspect is that the sensitivity of the line to Zeeman splitting, summarized by $w$ in Table~\ref{tab:observations}, does not correlate with the relative intensity of the line. This is of primary importance as many broadening effects correlate with the peak intensity of the line, such as the asymmetric perturbations associated with unresolved spatial intensity gradients \citep[e.g.,][]{Keppler_ea_2019, Boehler_ea_2021} or contamination from `back-side' emission \citep[e.g.,][]{Izquierdo_ea_2025}, so a clear correlation between line width and peak intensity would signify broadening mechanisms unrelated to Zeeman splitting. To demonstrate this, Figure~\ref{fig:app:broadening_sensitivity} plots the line width ratio (again, derived by the fit of a Gaussian line profile to the resulting spectrum) of the most sensitive hyperfine component to the least hyperfine component as we increase different model parameters which impact the line width. As can be seen, it is only the magnetic field strength, $|\mathbf{B}|$, which results in an increasing ratio, while all the others show minimal change in the ratio, despite all lines getting broader. The slight decrease observed here is due to the less sensitive line, $N=1-0$ $J=3/1-1/2$ $F=3/2-1/2$, having a substantially larger optical depth and so the impacts of non-Gaussianity due to optically thick cores reduces the ratio.

\begin{figure*}
    \centering
    \includegraphics[width=\textwidth]{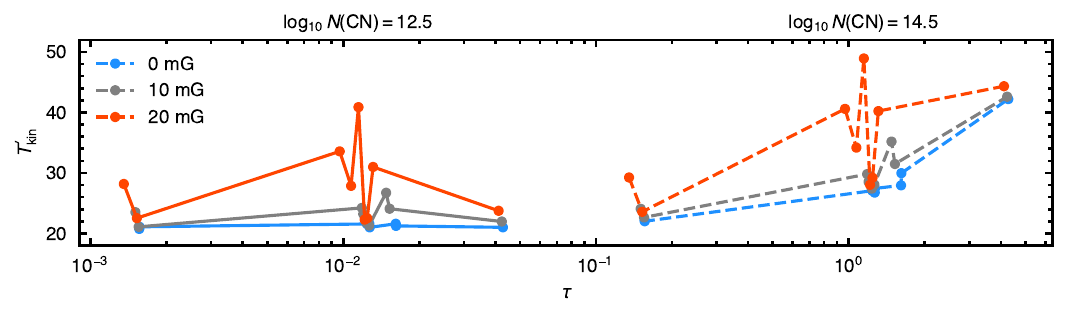}
    \caption{Plotting the temperature required to explain the observed line width for each hyperfine component when considering different underlying magnetic field strengths, shown by different colors, and a different CN column density, shown by different line styles. As the sensitivity of a line to Zeeman splitting is independent of the optical depth, lines with middling optical depths are observed to broaden more. The rise and fall in the effective kinetic temperature as a function of optical depth indicates that an unrealistic vertical temperature gradient would need to be invoked in order to explain the observed ratio of line widths without the use of Zeeman splitting.}
    \label{fig:app:temperature_tau_correlation}
\end{figure*}

One final consideration is the impact of the disk vertical structure which is not considered in the excitation analysis that assumes the emission arises from a homogeneous slab. As each of the hyperfine components has a different optical depth, it is possible that their emission arises from substantially different conditions which may lead to varying line profiles for each hyperfine component. As previously mentioned, the fact that the sensitivity to Zeeman splitting for each of the lines does not correlate with the relative intensity of the line suggests that this should not result in changes that could be misinterpreted as Zeeman splitting. To demonstrate that this is the case, Figure~\ref{fig:app:temperature_tau_correlation} plots the effective kinetic temperature for each line (the kinetic temperature one would infer based on the line width assuming that thermal broadening was the sole broadening mechanism) as a function of their optical depth for different magnetic field strengths (different colors) and CN column densities (different line styles). As can be seen, the increased Zeeman sensitivity of the lines with average optical depths lead to broader lines which would be interpreted as arising from warmer regions. Thus, assuming that $\tau$ is a proxy of the height above the midplane that the emission arises, one would need to invoke an unrealistic vertical temperature structure to explain the observed line broadening ratios, demonstrating that such an effect could not masquerade as Zeeman broadening.

\section{Model Fits}
\label{sec:app:model}

\begin{figure*}
    \centering
    \includegraphics[width=\textwidth]{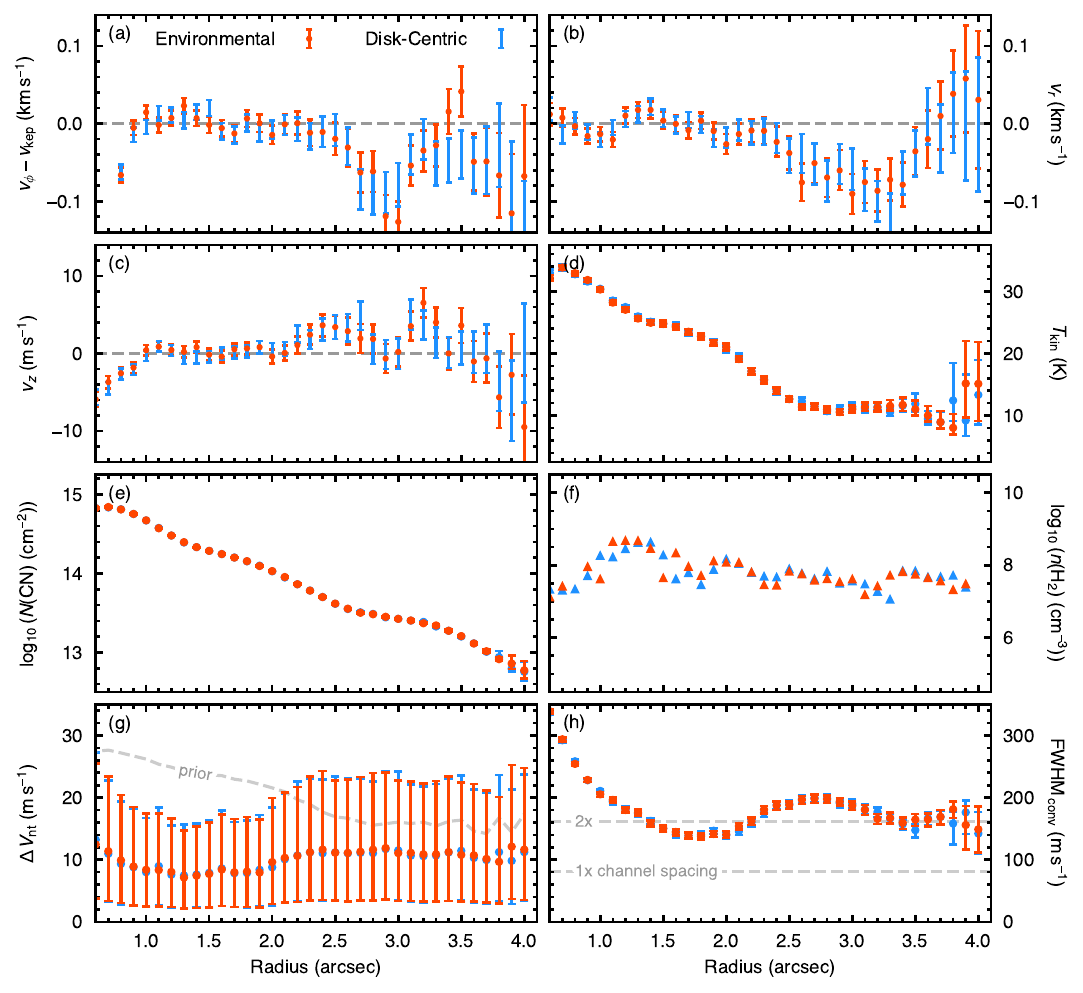}
    \caption{Excitation conditions derived from the fitting of the spectra in addition to the magnetic field properties plotted in Fig.~\ref{fig:B_theta_radial_profiles}. The blue and the red points show the disk-centric and environmental parameterizations of the magnetic field, respectively. For the local gas density, the arrows represent lower limits to this value -- essentially the critical density of the $N=1-0$ transition for the given temperature and column density. The gray dashed line for the non-thermal width component represents the $0.08~c_s$ prior, while the two horizontal lines in the convolution kernel panel show 1 and 2 times the channel spacing of $80~{\rm m\,s^{-1}}$.}
    \label{fig:app:excitation_conditions}
\end{figure*}

In this section we discuss the other parameters derived as part of the spectral analysis shown in Figure~\ref{fig:app:excitation_conditions}. As with Fig.~\ref{fig:B_theta_radial_profiles}, blue points denote the model with a disk-centric magnetic field morphology while the red points show the environmental field morphology. All error bars represent the 16th to 84th percentile range of the posterior distribution. It is clear (and expected) that the assumed field morphology should have no influence on any of these parameters.

\subsection{Excitation Analysis}
\label{sec:app:model:excitation}

Panels (a), (b) and (c) show the recovered azimuthally-averaged velocity profiles. To bring out the structure in the velocity profile, panel (a) shows the rotation velocity after subtracting a Keplerian velocity profile for a stellar mass of $0.82~M_{\odot}$ and source inclination of $5\fdg8$. There is negligible deviation from this Keplerian profile out to $2\farcs5$ (150~au) where a slowing of the rotation is observed, likely due to the onset of the pressure support from the exponential taper of the gas surface density \citep{Teague_ea_2022}. Interior to $1\arcsec$ there is a considerable deviation, however this is dominated by beam smearing effects \citep[e.g.,][]{Teague_ea_2018a}. There is no discernible radial velocity component inwards of $2\farcs5$, however around a deviation is inferred with a maximum velocity of $|v_r| \approx 100~{\rm m\,s^{-1}}$, as shown in panel (b). Unfortunately, given the close to face-on nature of the disk around TW Hya, we are unable to robustly differentiate inwards and outwards motions, although the clockwise motion suggested by the spiral arm would suggest an \emph{inwards} motion \citep{Teague_ea_2022}. As such, it is unclear whether this is a signature of a disk wind or another phenomena. No significant vertical motions are detected outwards of $1\arcsec$, although there is a large scatter. Interior to this a vertical component is required to reproduce the data, however this is likely due to beam smearing effects, as with $v_{\phi}$.

As shown in panel (d), the kinetic temperature is found to decrease radially, starting from above 30~K in the inner disk and dropping to around 10~K outside $2\farcs5$ where it plateaus. This appears to be close to the expected midplane temperature \citep{Macias_ea_2021}. The column density of CN also drops radially, shown in panel (e), ranging from just under $10^{15}~{\rm cm^{-2}}$ at $0\farcs5$, down to less than $10^{13}~{\rm cm^{-2}}$ beyond $3\farcs5$. Two subtle dips are perceptible in both the kinetic temperature and column density radial profiles at ${\sim}~1\farcs3$ (78~au) and ${\sim}~2\farcs7$ (162~au). Depressions at these radial locations was also reported in the CS (5-4) integrated flux profile in \citet{Teague_ea_2022} suggesting that these regions may be gaps in the gas surface density. 

\begin{figure*}
    \centering
    \includegraphics[width=\textwidth]{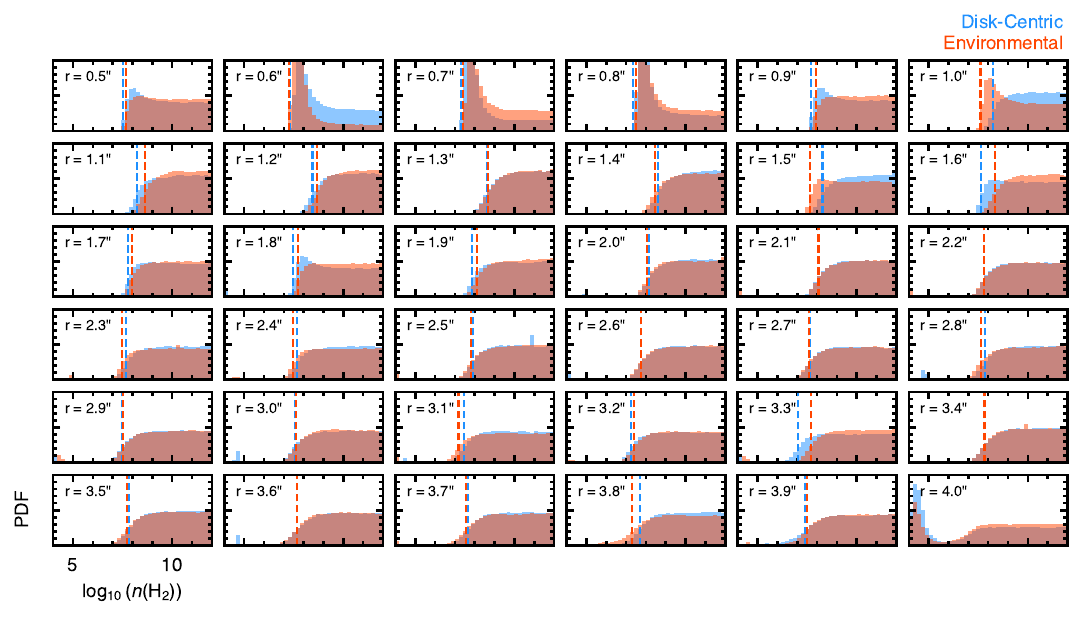}
    \caption{The posterior distributions for the collider density at each radius. The vertical dashed lines show the inferred lower limit to the density based on the median value of a Gaussian cumulative probability distribution function fit to the distribution.}
    \label{fig:app:nh2_posteriors}
\end{figure*}

The collider density, assumed to be solely H$_2$, is show in panel (f) as lower limits. These are the critical densities for the (1-0) transition for the given temperature and column density of CN. To derive this lower limit from the posterior distributions of $n({\rm H_2})$, shown in Fig.~\ref{fig:app:nh2_posteriors} for both field morphologies, we fit a scaled cumulative Gaussian distribution function and use the median of the Gaussian distribution as the lower limit to the density. These are shown as the dashed vertical lines in Fig.~\ref{fig:app:nh2_posteriors}. Interior to $1\arcsec$, the posteriors appear to be peaked around the critical density, however this is likely due to beam smearing effects perturbing the line profiles and thus we suggest that these are not robust density estimates.

Panels (g) and (h) show the systematic broadening mechanisms: non-thermal broadening and a Gaussian convolution of the final spectrum, respectively. The $1\sigma$ width of the prior on the non-thermal broadening based on the upper limts from \citet{Flaherty_ea_2018} are shown in (g). Across the bulk of the disk, the non-thermal broadening is consistent with zero, in agreement with previous searches for non-thermal broadening in this source \citep{Hughes_ea_2011, Teague_ea_2016, Flaherty_ea_2018}. Beyond ${\sim}~2\farcs5$, non-zero values are found needed to account for the extremely narrow thermal widths in these regions (a kinetic temperature of 10~K produces a thermal Doppler width of $80~{\rm m\,s^{-1}}$. With such narrow lines, the limited spectral resolution of the data will prohibit robust inferences about the line width, as evidenced by the magnetic field inferences shown in Fig.~\ref{fig:B_posteriors_disk} and \ref{fig:B_posteriors_env}.

\begin{figure*}
    \centering
    \includegraphics[width=\textwidth]{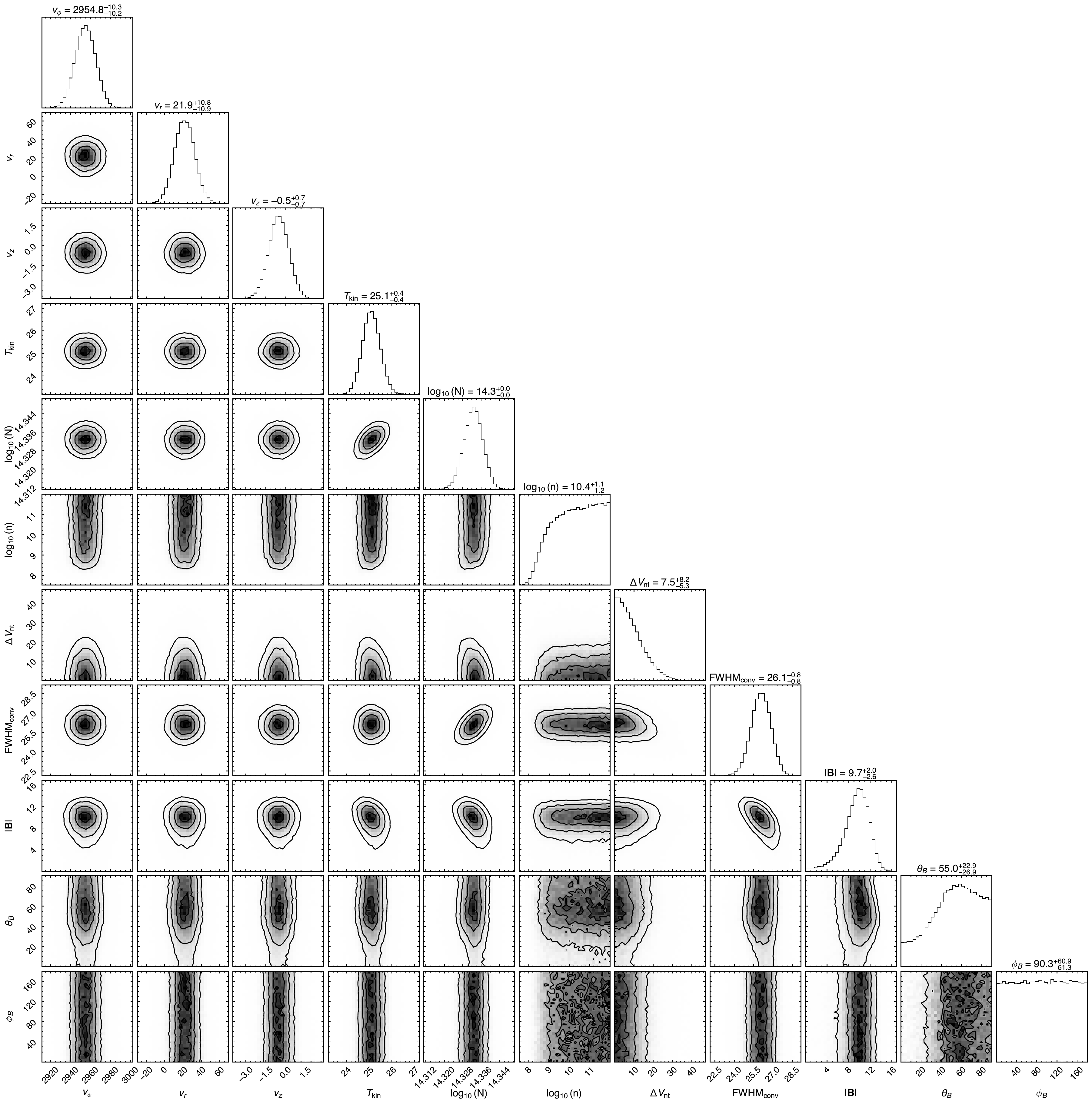}
    \caption{Example corner plot showing the posterior distributions and covariances for the disk-centric model at $1\farcs4$. No significant correlations are found between parameters.}
    \label{fig:cornerplot_disk}
\end{figure*}

Figure~\ref{fig:cornerplot_disk} shows the marginalized posterior distributions and covariances for all the parameters of the model for a disk-centric magnetic field optimized for $1\farcs4$ annulus in the disk. No significant correlations are found, however a slight covariance is observed between ${\rm FWHM_{conv}}$ and $|\mathbf{B}|$ which is unsurprising given that both influence the width of the resulting line.

\citet{Yoshida_ea_2024} performed a similar non-LTE analysis as described here and found consistent excitation conditions across the radial region where these two analyses overlapped: $0\farcs6$ and $2\arcsec$. \citet{Yoshida_ea_2024} used these conditions to suggest an emission height for CN to arise from $z \, / \, r 
\sim 0.2 - 0.3$, consistent with the previous estimate from \citet{Teague_Loomis_2020}.

\subsection{Zeeman Sensitivities}
\label{sec:app:model:zeeman}

\begin{figure*}
    \centering
    \includegraphics[width=\textwidth]{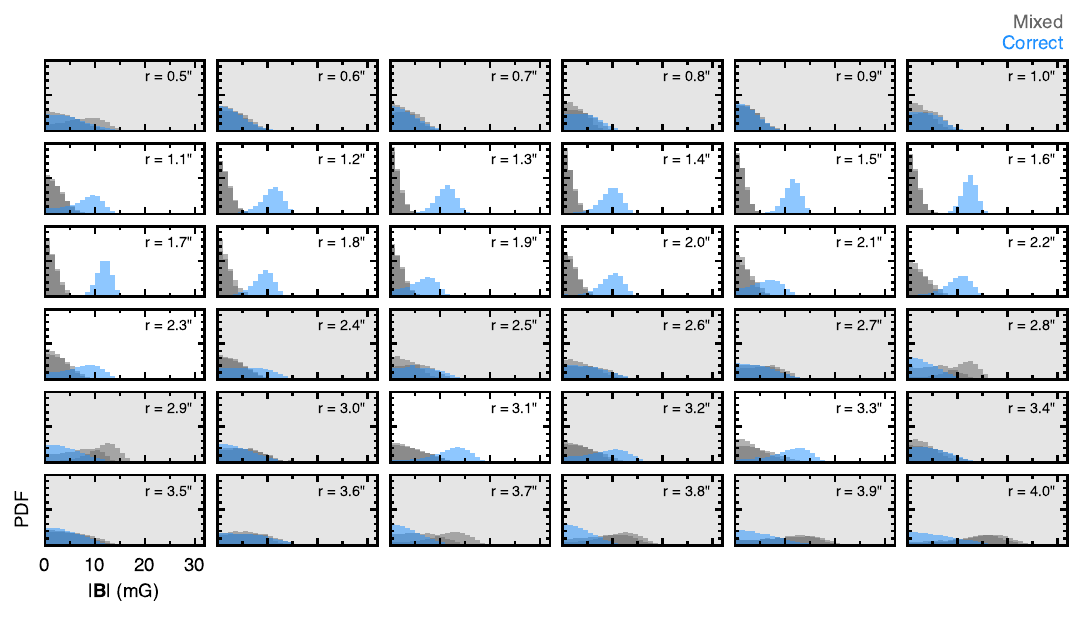}
    \caption{A comparison of the posteriors probability distribution for the magnetic field strength at each radial location. Each panel shows the posteriors for the disk-centric parameterization in blue while the two gray posteriors are for the same model, but when the Zeeman sensitivities of the two most sensitive hyperfine components are swapped with the two least sensitive hyperfine components. If the `correct' mapping of sensitivities finds a higher magnetic field strength than the `mixed' mapping then this is strong evidence that it is truly Zeeman broadening that is observed. Panels where the medians of the distributions are separated by less than the quadrature sum of the standard deviations of the distributions are colored in gray. This appears to be the case for separations ${\gtrsim}2.5\arcsec$. Note that for most radii, the two `randomized' models have posteriors which overlap and cannot be distinguished.}
    \label{fig:B_posteriors_disk}
\end{figure*}

\begin{figure*}
    \centering
    \includegraphics[width=\textwidth]{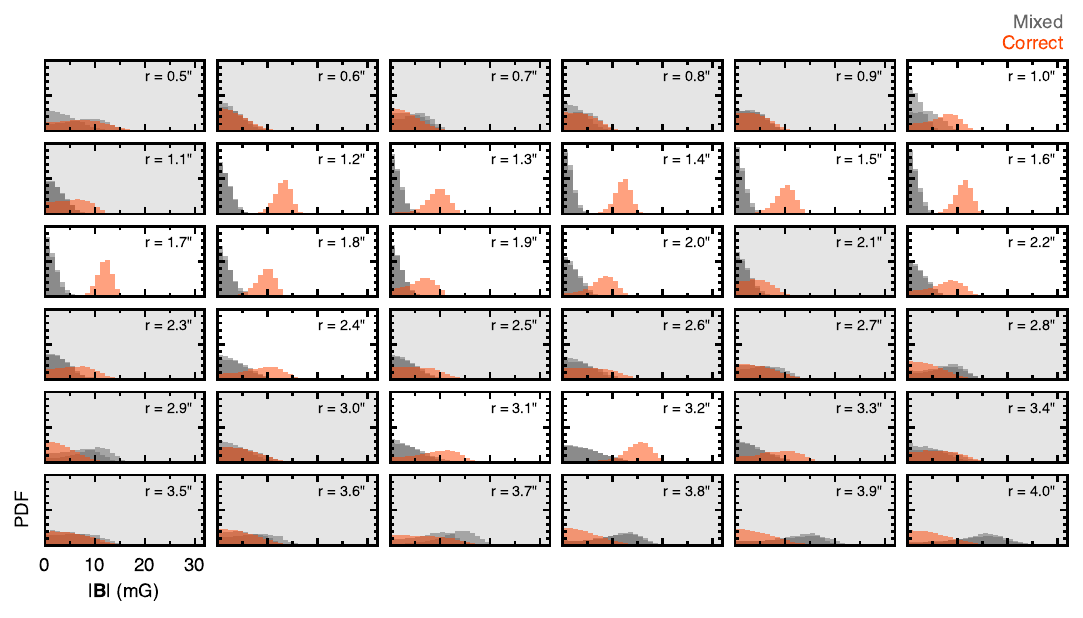}
    \caption{As Figure~\ref{fig:B_posteriors_disk} but for the environmental field morphology.}
    \label{fig:B_posteriors_env}
\end{figure*}

Figures~\ref{fig:B_posteriors_disk} and \ref{fig:B_posteriors_env} compare the posterior distributions for the magnetic field strength when the line-specific sensitives to the Zeeman effect are correct, in color, and mixed up in gray. The two gray models are when the most sensitive transitions, $J=3/2 - 1/2$ $F = 3/2 - 1/2$ and $J = 1/2 - 1/2$ $F = 3/2 - 3/2$, are swapped with the least sensitive transitions, $J=3/2 - 1/2$ $F = 1/2 - 1/2$ and $J=3/2 - 1/2$ $F = 3/2 - 3/2$, and overlap for most radii. Each panel shows a different radial bin and those with a gray background are where the two distributions are not statistically distinct. For this measure we calculate the posterior distance as the difference between the two posterior medians and require this to be larger than the quadrature sum of the two posterior widths, taken as the 16th to 84th percentile range. As the posteriors for the two `random' models are extremely similar for most radii, we combine them and treat them as a single posterior distribution for this calculation.

Over most of the disk the `correct' energy level structure posteriors appear to peak at larger magnetic field strengths than for the `mixed' structures, as would be expected for Zeeman splitting, however the large widths of the posteriors in the outer disk, $\gtrsim 2\farcs5$, preclude a robust differentiation between the posteriors.

\end{appendix}

\bibliography{main}{}
\bibliographystyle{aasjournal}

\end{document}